# *Multicores-periphery structure in networks*

Bowen Yan and Jianxi Luo

Engineering Product Development Pillar and SUTD-MIT International Design Centre,
Singapore University of Technology and Design, Singapore, 487372

(bowen_yan@sutd.edu.sg and luo@sutd.edu.sg)

## Abstract

Many real-world networks exhibit a *multicores-periphery structure*, with densely connected vertices in *multiple cores* surrounded by a general periphery of sparsely connected vertices. Identification of the multicores-periphery structure can provide a new lens to understand the structures and functions of various real-world networks. This paper defines the multicores-periphery structure and introduces an algorithm to identify the optimal partition of multiple cores and the periphery in general networks. We demonstrate the performance of our algorithm by applying it to a well-known social network and a patent technology network, which are best characterized by the multicores-periphery structure. The analyses also reveal the differences between our multicores-periphery detection algorithm and two state-of-the-art algorithms for detecting the single core-periphery structure and community structure.

# 1 Introduction

Many real-world systems can be represented as networks, for instance, social networks, technological networks, information networks, and biological networks. In the past two decades, various algorithms have been developed to explore the structures of real-world networks, which may reveal the properties and functions of the respective networks (Newman, 2003; Strogatz, 2001). A particular and popular strand of network analyses has focused on detecting meso-scale structures, such as *communities* (or clusters) in networks. Vertices in the same community are more cohesively connected to each other than those in different communities (Fortunato, 2010).

The *core-periphery structure* is an alternative meso-scale structure that has been discovered in many real-world networks, such as social networks, transportation networks and the World Wide Web (Borgatti & Everett, 2000; Csermely et al., 2013; Rombach et al., 2014). A network characterized by the core-periphery structure exhibits some sort of *core*, in which vertices are densely connected, and a periphery, in which vertices are only sparsely connected. Both community and core-periphery structures have important implications on the functions in the networks that embed them (Zhang et al., 2015). For instance, in communication networks, dense connections in a dense community or core may lead to efficient information flow or synchronization among vertices in the same community or core (Wasserman & Faust, 1994; Xu & Chen, 2005). In social networks, persons in the densely connected core might be more influential or powerful than those in the periphery.

However, real-world networks can exhibit ***multiple*** cores, each of which contains vertices that are only densely connected to each other within the respective cores, together with the periphery, in which vertices are only sparsely connected in general. For example, in a social network people may be cohesively connected in different sub-groups; meanwhile there are always people who are only loosely connected to any of the sub-groups and other people in general. A city may have multiple dense centers (i.e., cores) for different urban functions, and a general sparse suburb (i.e., periphery) surrounding them. Rombach et al. (2014) observed two cores in London's underground railway network. Zhang et al. (Zhang et al., 2015) visually identified two cores in the network of hyperlinks between political blogs, leaving those generally loosely connected blogs in the periphery. In our earlier analysis of the structure of a weighted network of patent technology classes that represent the total technology space (Yan & Luo, 2017), we vaguely observed *several strong cores* which contain technology classes that are strongly and cohesively related to one another, and the periphery consists of all outlying and weakly-connected technology classes.

These networks exhibit a meso-scale structure in common, i.e., *multiple cores*, each of which contains densely connected vertices, surrounded by the periphery, which contains the sparsely connected vertices. We refer to this structure as a “multicores-periphery structure”. Figure 1 illustrates the fundamental differences between the newly defined *multicores-periphery* structure and the well-known community and core-periphery structures. Moreover, various studies have attempted to identify the periphery closely connected to a clique with a maximum or required density, and resulted in a structure of multiple sets of dense cores and their own affiliated peripheries (Bruckner et al., 2015; Everett & Borgatti, 2000; Yang & Leskovec, 2014). The multicores-periphery structure defined in this paper differs from those with “multiple cores, each with its own periphery (Borgatti & Everett, 2000)” in that it contains only one general periphery, which hosts all the vertices with a generally weak connectivity. The peripheral vertices are basically outliers and not affiliated with any core. Mathematical formulas in chapter 3 manifest the definition here.

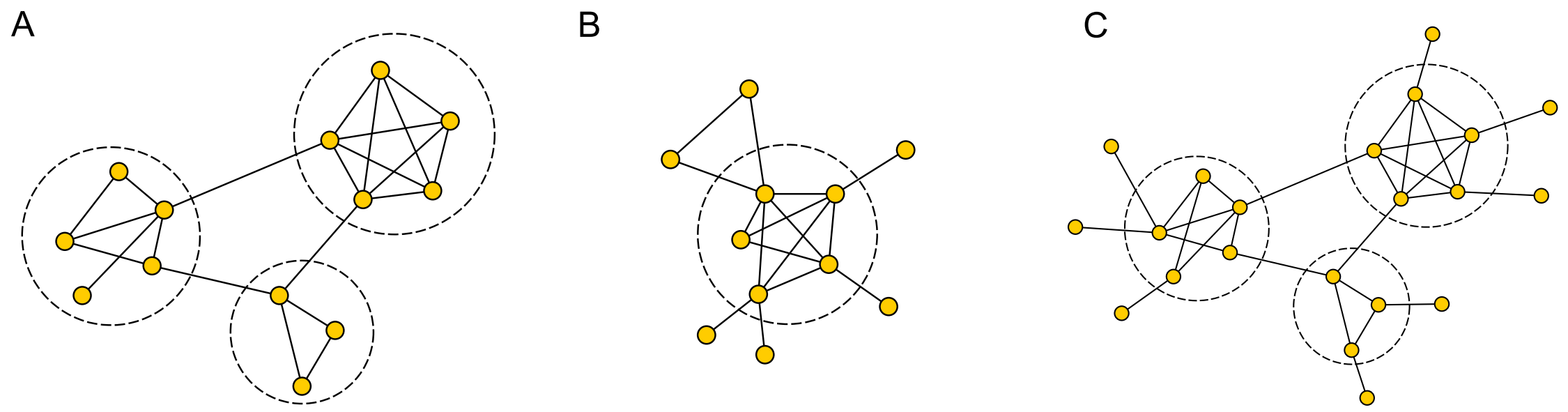

Fig. 1. Examples of meso-scale network structures: (A) community; (B) core-periphery; (C) multicores-periphery.

In this paper, we focus on the quantitative identification of the *multicores–periphery* structure in real-world networks. The detection of the multicores-periphery structure has potential uses, particularly for the networks that innately have multiple cohesive cores and many generally outlying elements that are not affiliated with any core. Different cores and the general periphery may play different functions or roles in a network, whereas such roles despite their differences might be generally more important than the peripheral ones. For instance, a city may consist of multiple core areas for governance, business, shopping, entertainment, and education functions, and the peripheral suburb area for residential function. Simultaneously distinguishing different cores and the general sparse periphery may shed new light on the functions in networks that are inherently governed by a multicores-periphery structure.

To the best of our knowledge, there is still no method to directly detect the *multicores-periphery* structure, i.e., *multiple cores plus a general periphery that surround the cores*, in networks. The existing core-periphery network literature has focused on either dichotomous or continuous division of the network into a single core and a periphery (Borgatti & Everett, 2000; Csermely et al., 2013; Holme, 2005; Rombach et al., 2014). Although existing community detection algorithms are able to locate multiple similarly defined cohesive communities, the periphery is not considered or differentiated (Fortunato, 2010). The methods to detect cliques, *k*-scores or *n*-plexes may find multiple local cohesive sub-groups with pre-determine densities, but do not concern the separation and distinction from the remaining network (Brandes & Erlebach, 2005). Prior studies have explored ways to identify the peripheries that are affiliated to specific dense cores defined as cliques with pre-determined densities, but such analyses result in multiple peripheries only affiliated to their own cliques (Bruckner et al., 2015; Everett & Borgatti, 2000). In contrast, following the multicores-periphery structure of our interest and definition, the dense cores are not required to be cliques or have specific densities, and there is only one general periphery rather than multiple peripheries.

Note that, nested uses of existing core-periphery or community detection algorithms may lead to a structure that has multiple "cores". For instance, one can use existing algorithms to detect communities first, and then decompose each community into a core and a periphery. However, this would result in multiple peripheries with community identities, which violate our definition that there is only a general sparse periphery. Alternatively, one can first detect a single core and periphery division and then decompose the core into sub-communities. This sequence may yield stronger connections between the resultant "sub-cores" than normal, because an optimal core identified by a core-periphery algorithm (Borgatti & Everett, 2000; Holme, 2005) would be internally cohesive. Especially, when the core is extremely dense or cohesive, i.e., it is a clique with maximum density (Brandes & Erlebach, 2005; Everett & Borgatti, 2000; Yang & Leskovec, 2014), it would be theoretically incorrect to arbitrarily detect communities in the non-decomposable core. In general, community and single core-periphery detection algorithms, as well as clique detection algorithms, were developed for other structures. The nested use of them would be a methodological compromise, and by nature unable to capture the exact meaning of the multicores-periphery structure that we define here.

In this paper, we present a method to directly identify the optimal partition of a network that can have multiple dense cores and a general sparse periphery, i.e., the multicores-periphery structure. Our method defines a single statistical measure that characterizes the distinction between the multiple cores and the periphery in terms of connection density, and examines the value given by every possible network partition in the hierarchical dendrogram of the network. The optimal partition emerges when the value of the measure characterizing the *multicores-periphery* structure is maximized. This method is generally applicable to any type of networks. We demonstrate that the method successfully identifies *multicores-periphery* structures in a social network and in a technology space network. Both networks are weighted.

# 2 Core-periphery structures in networks

The comprehensive review of the core-periphery structure in networks by Csermely et al. (2013) revealed various definitions and types of core-periphery structures in biological, social-economic and technological network literatures. Despite the variety of definitions, they are generally consistent in the intuition that vertices in the core are densely connected and vertices at the periphery are sparsely connected. Some believe that core vertices should also be well-connected to the periphery (Borgatti & Everett, 2000; Csermely et al., 2013; Rombach et al., 2014). Some others state that core vertices should be both densely connected and central to the network (Holme, 2005).

Borgatti and Everett (2000) proposed an association function on how far a given network deviates from a comparable network with an ideal core-periphery structure and minimized this function to find the best core-periphery partition of the network. In their ideally defined core-periphery structure, core vertices are fully connected to each other and to the peripheral vertices, but the peripheral vertices are not connected to each other. In a similar spirit, Zhang et al. (Zhang et al., 2015) proposed an algorithm to identify the core-periphery structure by fitting a stochastic block model to empirical network data using a maximum likelihood method. Borgatti and Everett (2000) also defined a continuous core-periphery division, in which every vertex is assigned a *coreness* value that quantifies its qualification to be in the core. Rombach et al. (2014) presented a more flexible model to determine the fraction and sharpness of the core-periphery division. They employed a *transition function* to assign a *core score* to each vertex and maximized the quality function of a core by simulated annealing.

Other methods have defined and detected core vertices that are also central in the network, in addition to being well connected to each other (Holme, 2005; Shanahan & Wildie, 2012; Silva et al., 2008). For example, Holme (2005) assumed that core vertices should have a high closeness centrality, i.e., a short average distance from the rest of the vertices, and proposed a core-periphery coefficient to measure the extent to each an empirical network exhibits a clear-cut single core-periphery dichotomy. In particular, he defined the single core as the *k*-core with the maximal closeness centrality among all *k*-cores. Silva et al. (2008) considered both closeness centrality and community modularity (Newman, 2004) to identify the core-periphery structure of metabolic networks. They showed that closeness centrality provides better results than degree and betweenness centralities for identifying core vertices. Della Rossa et al. (2013) defined a core-periphery profile for the network along with a *coreness* value for each vertex that were calculated by following the order of the degree centrality of each vertex based on a random walk model. These methods generally aimed to identify a single core in the core-periphery structure.

Some prior studies borrowed the concepts of cliques or *k*-cores to define and detect multiple cores. For example, to study protein interaction networks, Bruckner et al. (Bruckner et al., 2015) detected multiple cores defined as complete cliques, in each of which all proteins must interact with each other. Likewise, Everett and Borgatti (Everett & Borgatti, 2000) additionally discussed the relaxed cases where the cores are *k*-cores, in which all vertices have a degree of at least *k*. *K*-cores are relaxed cliques, and cliques are extreme cases of *k*-scores with maximum *k*. The definitions of cliques and *k*-scores have constrained the analyses only possible for unweighted networks. In addition, the detection of cliques or *k*-cores requires pre-determined local densities, does not take into account the global network structure and the separation and relative distinction from the remaining network (Brandes & Erlebach, 2005), and is unable to guarantee an optimal network partition into multiple cores and the periphery.

Indeed, Everett and Borgatti (Everett & Borgatti, 2000) and Bruckner et al. (Bruckner et al., 2015) primarily aimed to identify the separate peripheries that were most closely connected to respective cliques. Yang and Leskovec (Yang & Leskovec, 2014) also detected multiple separate peripheries, but their peripheries share affiliations with dense cores. Their method does not rely on the rigid clique and cohesion concepts, but took the dense overlaps of multiple ground-truth functional (instead of structural) communities as cores, and the non-overlapping regions of the functional communities as peripheries. Each vertex must have one or multiple community identities. In general, these methods identifying multiple peripheries of dense cores are not in line with our definition that emphasizes one single generally sparse periphery and its separation from the cores.

In general, prior research on core-periphery structures focused on either continuous or dichotomous division of a network into a single core and a periphery. The studies that revealed multiple cores either used rigid definitions of cliques with maximum or pre-determined densities, which do not concern the overall network structure and have limited applicability, or focused on affiliating multiple peripheries to their most close cores, apart from our definition of the multicores-periphery structure which considers all the vertices outside different cores belong to a general periphery. There is still no quantitative method to directly identify the *multicores-periphery* structure, i.e., the partition of a network into multiple cores surrounded by a general sparse periphery, which characterizes many real-world networks.

# 3 Finding the optimal partition of multiple cores plus periphery

Herein, we introduce a method to detect the optimal multicores-periphery partition of a given network as the partition in its hierarchical dendrogram that provides the largest distinction between cores and the periphery in terms of connection densities. The first step is to create a dendrogram of the network, using a hierarchical clustering algorithm. In the dendrogram, more closely connected vertices are joined by shorter and lower branches than those more distantly connected vertices. The dendrogram defines a series of network partitions from the bottom where every vertex is stand-alone, to the top where all vertices belong to one single cluster. In each partition, some vertices are in clusters (i.e., potential cores) and some other vertices stand alone outside any cluster.

When creating the dendrogram, the average linkage clustering algorithm is used to determine the distance between each pair of temporary clusters of vertices (in the dendrogram) as the mean of all pairwise distances between vertices in both clusters (Wasserman & Faust, 1994). Compared to alternative clustering methods (for instance, the single-linkage clustering method that uses the minimum distance and the complete-linkage clustering method that uses the maximum distance between vertices in different clusters), the average-linkage clustering method prioritizes the clusters with the strongest cohesion to merge first, so that the merged cluster would have the highest internal connection density, because density is correlated with average link weight in a cluster. Such a property is desirable because the definition of the cores-periphery structure emphasizes high link density within each core.

Among all the partitions in the dendrogram, the optimal cores-periphery partition should have the largest possible distinction between cores and the periphery in terms of respective connection densities. Such a distinction for a partition is calculated as the ratio of the density of connections of vertices in all clusters (i.e., potential cores) over the density of connections of the vertices that are not assigned to any cluster (i.e., potential periphery). The densities of connections in the cores and in the periphery are calculated as,

$$\text{density}_{cores} = \frac{C}{\sum_{i=1}^{k} \frac{n_i(n_i-1)}{2}} \quad (1)$$

$$\text{density}_{periphery} = \frac{P}{\frac{m(m-1)}{2} + m\sum_{i=1}^{k} n_i} \quad (2)$$

where $n_i$ is the number of vertices in cluster $i$, for $i$=1 to $k$ clusters; $m$ is the number of vertices at the periphery; $C$ is the sum of weights of all the connections of core vertices within cores; and $P$ is the sum of weights of all the connections of peripheral vertices outside cores. Then, we define the following *cores-periphery* ratio (note the plural "*cores*" in the term) to measure the degree to which the partition fulfills a multicores-periphery structure,

$$r = \frac{\text{density}_{cores}}{\text{density}_{periphery}} \quad (3)$$

We further compared the empirically observed *cores-periphery* ratio $r$ (Equation 3) to those expected by chance using randomized networks. In the randomized networks, all edges between all vertices in the original network were switched using a Monte Carlo method, while preserving the weighted degree distributions of each vertex, in order to ensure that the observed and randomized networks have the same macro network structures so that they can be compared. To compare the observed $r$ with those of the randomized but comparable networks, we calculated a $z$-score for each network partition,

$$z = \frac{r_{\text{obs}} - r_{\text{rand}}}{\sigma} \tag{4}$$

where $r_{\text{obs}}$ is calculated from the partition of the empirical network and $r_{\text{rand}}$ and $\sigma$ are the mean and standard deviation, respectively, of the $r$ ratios calculated of a sufficiently large ensemble of randomized networks based on the same partition as that of the original network. This normalized measure describes the extent to which the network partition exhibits a cores-periphery structure relative to chance. Therefore, the partition that provides the maximal $z$ score is the optimal cores-periphery partition of the network.

Note that, the algorithm does not impose the assumption of finding more than one core, and may still allow a single core-periphery structure to be detected as optimal. That is, the optimal partition may emerge to have either a single or multiple cores, depending on the innate structure of the examined network. In contrast, the existing core-periphery algorithms impose the assumption of a single core for partitioning, and ignore the fact that many real-world networks are inherently organized and governed by a multicores-periphery structure.

Figure 2 illustrates the procedure of finding the optimal *multicores-periphery* partition in an example network. In brief, we first generate the dendrogram of a network, compute the $z$-scores, i.e., normalized cores-periphery ratios, for all possible partitions in the dendrogram, and then search for the partition with the highest $z$-score value. In that optimal partition, the clusters are the cores and all the vertices outside clusters belong to the periphery.

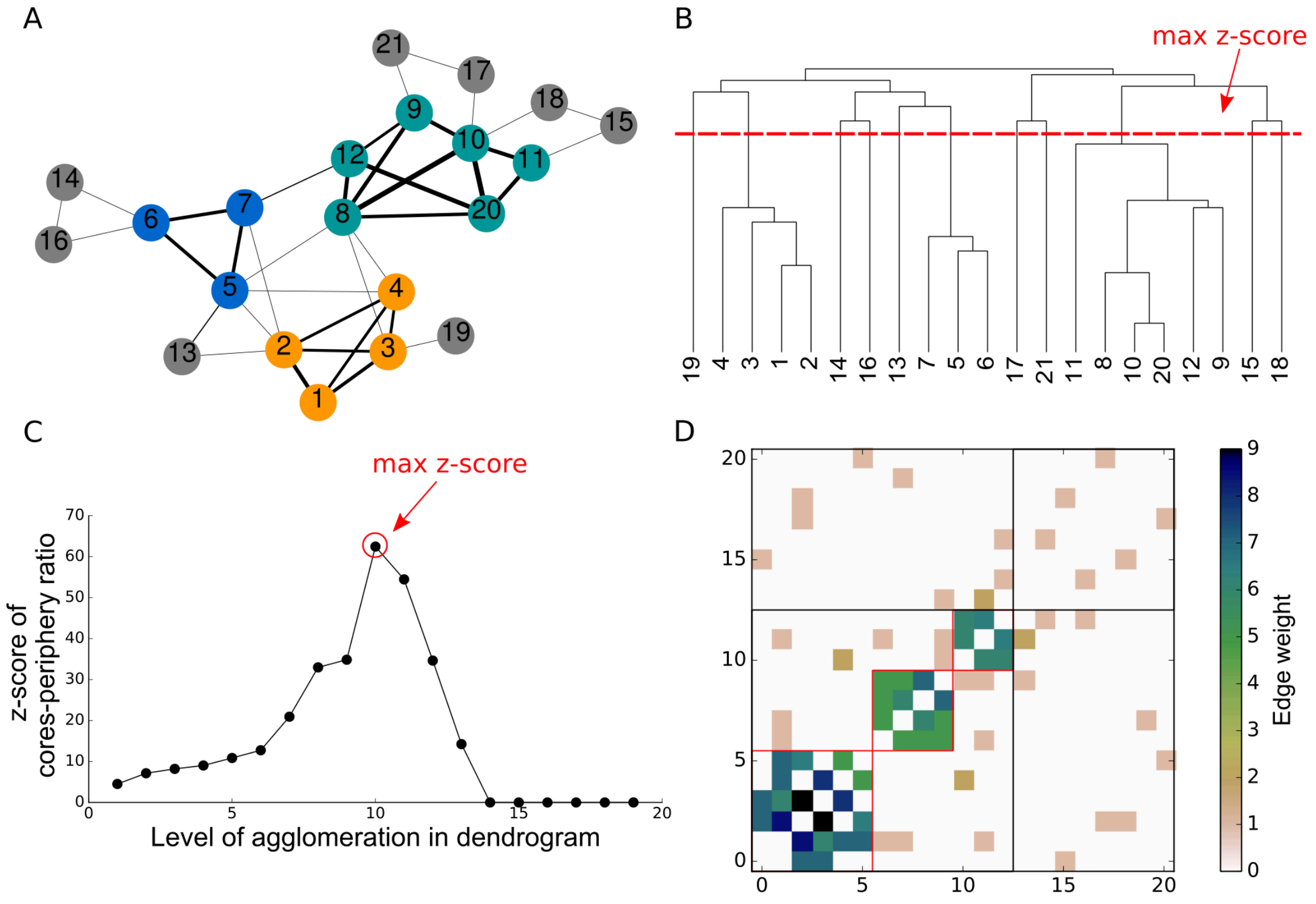

Fig. 2. The procedure of finding the optimal multicores-periphery partition in a sample network. (A) The original sample network of 21 vertices; (B) the dendrogram of the original network; (C) *z*-scores (calculated using ensembles of one million randomized networks) of different partitions; (D) matrix representation of the optimal multicores-periphery partition given by the maximal *z*-score.

The exemplar *z*-score curve in Figure 2(C) has only one peak, suggesting a single maximum *z*-score. However, the curve might display a plateau given a range of partitions for a maximum, or multiple peaks for equal maxima. In such cases, despite being unlikely, one may utilize the domain knowledge about the network to assess and compare the corresponding partitions giving the same maximal *z*-scores, in order to select the most reasonable or meaningful one(s).

In addition, given the structure of the algorithm described above, one can estimate its time complexity when applied to a real-world network. The first step of the algorithm is fundamentally the same as the average linkage clustering algorithm, so the time complexity is $O(n^2 \log n)$. The core of the second step is the generation of a randomized network, which means a time complexity $O(n^2)$. Understanding of the time complexity of our method will be useful to guide practitioners on the computational resources required when they apply the algorithm to real-world networks.

In the next section, we will demonstrate and test this method by applying it to detecting the *multicores-periphery* structures embedded in two distinct types of real-world networks, including a well-known social network and a network of patent technology classes that represent the technology space.

# 4 Multicores-periphery structure in real-world networks

Both real-world networks are undirected and weighted. Table 1 reports their original data sources, number of vertices and edges, maximum modularity score ($Q$) from the optimal network community partition given by the widely accepted Louvain algorithm (Blondel et al., 2008) and the maximal *cores-periphery* ratio ($z$-score) from our algorithm. In calculating the $z$-score for each candidate network partition, one million random networks are used. For each of the two real-world networks, we compare and analyze the optimal *multicores-periphery* structure identified by our method, the *core-periphery* structure detected by the algorithm of Borgatti and Everett (2000), and the optimal *community structure* identified by the Louvain community detection algorithm (Blondel et al., 2008).

**Table 1. Properties of real-world networks**

| Networks | Vertices | Edges | Max $Q$ | Max $z$ |
|---|---|---|---|---|
| Zachary's karate club network (Zachary, 1977) | 34 | 78 | 0.444 | 33.321 |
| Patent technology network (Yan & Luo, 2016, 2017) | 121 | 7120 | 0.251 | 85.547 |

## *4.1 Zachary's karate club network*

Zachary's karate club network is a social network of karate club members in the United States. This empirical network is well-known, as it has been popularly used as a benchmark case for testing different community detection algorithms (Fortunato, 2010). The network consists of 34 vertices that represent club members and 78 edges that represent social relationships among them. The weight of an edge between two club members is the count of social activities that both members attended together outside of the club, for instance, going together to a bar near the university campus. The karate club network was observed in the

period of 1970 and 1972.

At the time, two key persons in the network, the club president, John (vertex #34 in Figure 3(A)), and the instructor, Mr. Hi (vertex #1 in Figure 3(A)), had a conflict, which resulted in the separation of their respective social groups. Based on Borgatti and Everett's core-periphery detection method, one single core is found that contains both John and Hi (Table A1 in Appendices). Their single *core-periphery* partition does not reflect the conflict between John and Hi, and does not distinguish the separate social groups of these two persons. The Louvain community detection method places John and Hi in two separate large communities (Table A1), but does not differentiate the peripheral persons who are generally not social and stand neutral in the conflict between John and Hi. In contrast, in our optimal multicores-periphery partition, John and Hi belong to the two largest cores, together with a few small cores, which are surrounded by several generally non-social and insignificant individuals at the periphery (see Figure 3(A) and Table A1). The multicores-periphery structure provides the most systematic and nuanced characterization of the relative network positions and roles of different individuals in this social network.

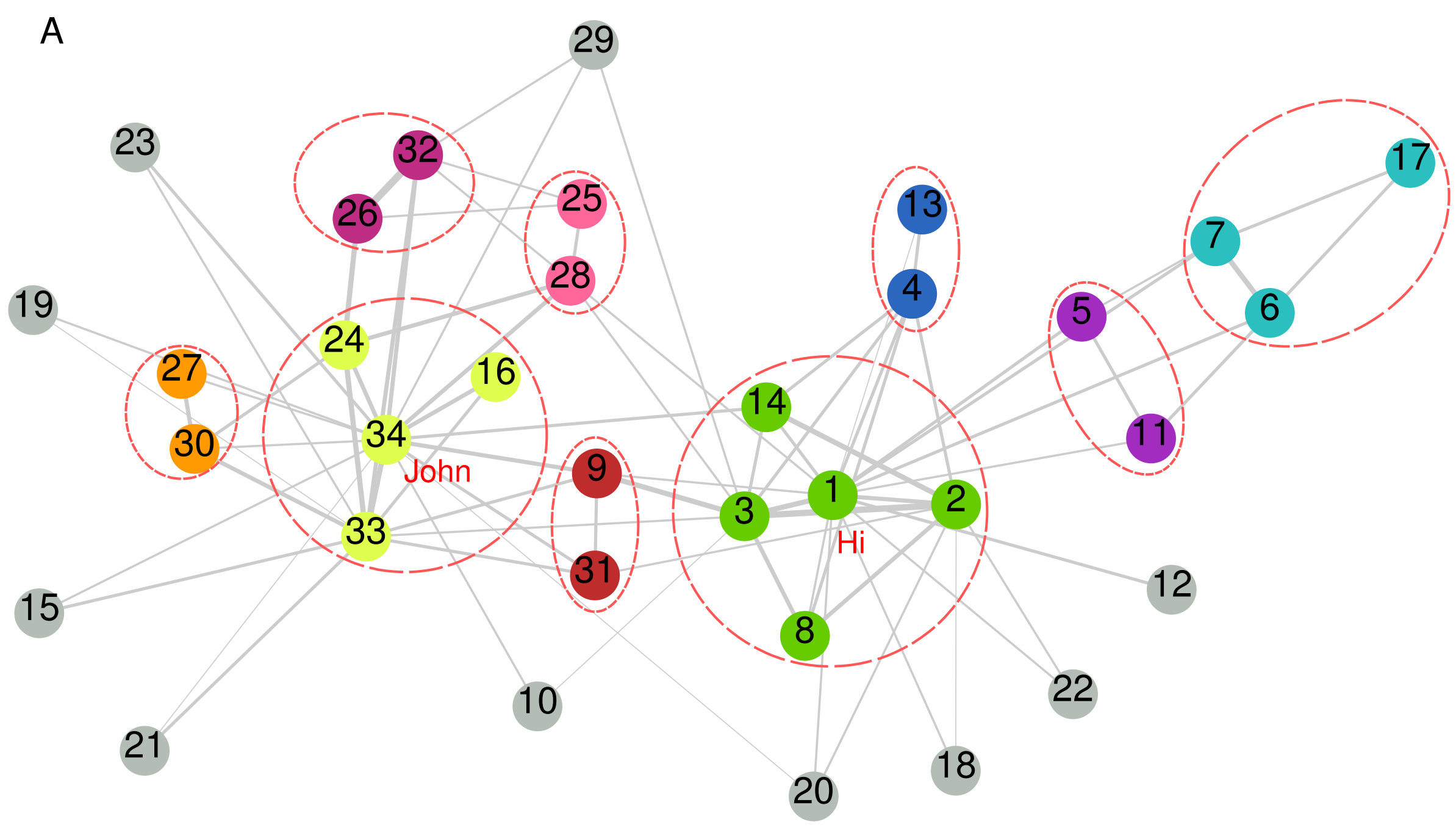

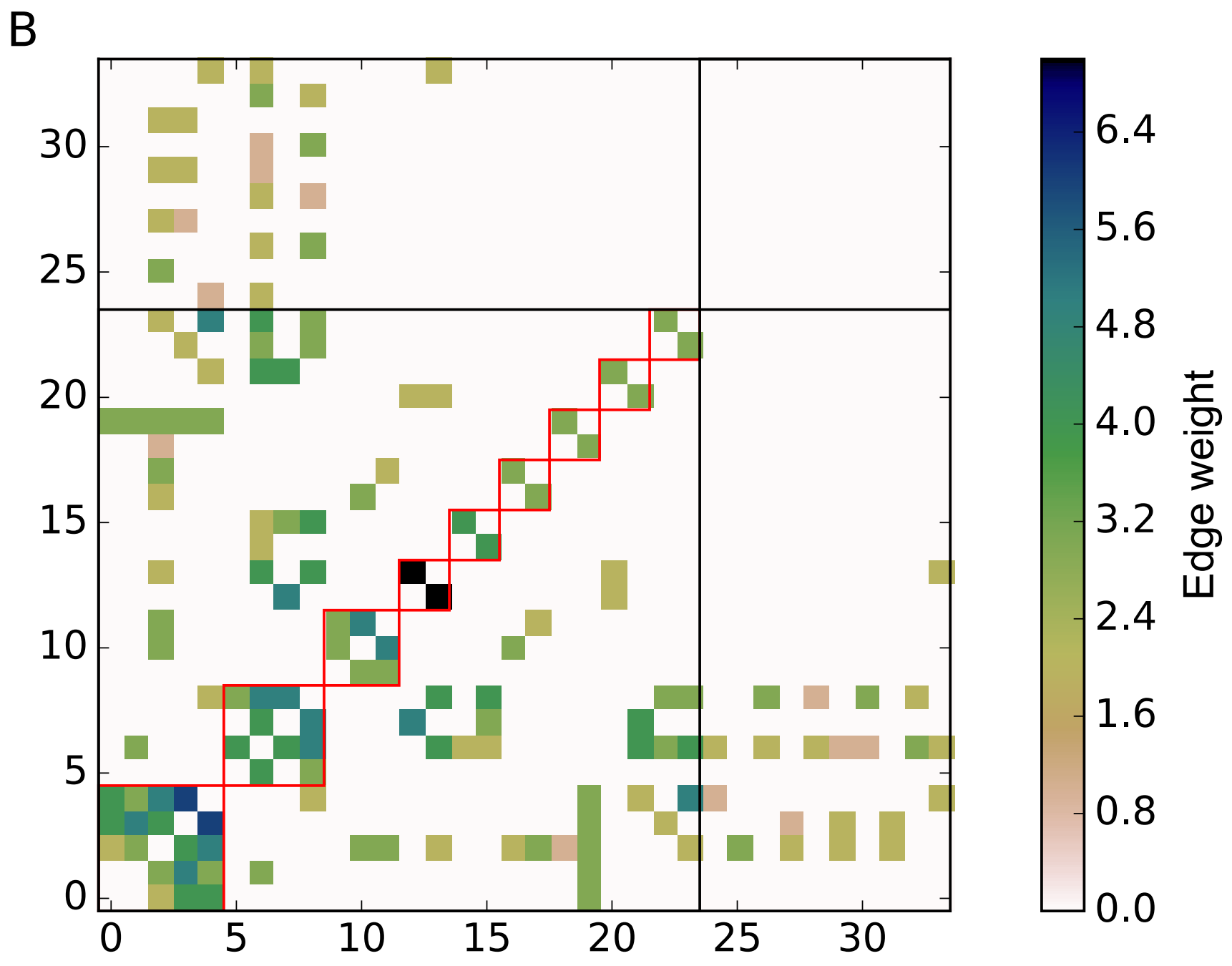

Fig. 3. Multicores-periphery structure of (A) Zachary's karate club network and (B) its matrix representation. In the network, each red circle encapsulates a core and the vertex colors are the same in the core and are different across cores; the peripheral vertices are in grey.

It is worth to note that nested core-periphery and community detections would not yield our partition that is holistically optimized according to the definition of the multicores-periphery structure. For example, the core where John is, including vertices #16, #24, #33 and #34, contains vertices #33 and #34 from the single core and vertices #16 and #24 from the periphery in the partition by Borgatti and Everett's core-periphery algorithm. That is, if one starts with the core-periphery algorithm to find the single core and then decomposes it to obtain more cores, he will never get the core of John because vertices #16 and #24 would never be located in any core. Likewise, the community detection result by the Louvain method separates the four vertices (#16, #24, #33 and #34) of the core of John in two communities. #24 belongs to one community, while #16, #33 and #34 belong to another. Therefore, if one starts with the community detection and then divides each community into a single core and periphery, he will never get the core of John because it is impossible for vertex #24 to join vertices #16, #33 and #34 in one core.

### *4.2 Patent technology network*

The patent technology network was constructed by Yan and Luo (Yan & Luo, 2017). Vertices are 3-digit patent technology classes defined by the International Patent Classification (IPC) system and represent different types of technologies, i.e., different technological areas in the total technology space. The strength of the edges represents the knowledge relatedness between pairs of IPC technology classes, and quantified as the Jaccard Index (Jaccard, 1901), i.e., the number of shared references of the patents in a pair of patent classes normalized by the number of unique references of the patents in either class. The edge values indicate the relatedness of the knowledge bases of the technologies represented by corresponding patent classes. To calculate the values of all edges, all patent records in the United States Patent and Trademark Office (USPTO) database from 1976 to 2010 were used.

For such a network of technologies connected according to their knowledge relatedness, an informative and meaningful structural partition must identify and reveal the groups of technologies that are densely related to each other within groups, but loosely related across groups. Applying Borgatti and Everett's core-periphery algorithm, we find 85 technologies in the single core and the rest in the periphery of the technology network (Table B2 in Appendices). The core contains (and thus does not discern) many distinct types of technologies, such as information technologies (e.g., G11, information storage; G05, controlling & regulating; G06, computing; G08, signaling), metallurgy-related technologies (B22, casting & metallurgy; C21, metallurgy of iron; and C22, metallurgy of non-ferrous metals or alloys), and food processing technologies (A21, baking; and A23, food processing). A meaningful partition needs to differentiate and separate such clearly distinct types of technologies into different groups.

In contrast, the optimal partition from our method identifies 27 cores and a general periphery, revealing a more nuanced and meaningful structure (Figure 4 and Table B2). For example, the metallurgy-related technologies (B22, C21, and C22), food processing technologies (A21 and A23), and information technologies (G05, G06, G08, and G11) now belong to different cores. Our optimal partition also reveals the meaningful cores that represent non-metal materials processing (B05, B28, B29, B32, C04, C08, and C09), biomedical technologies (A01, C12, A61, and C07), engine-related technologies (F01, F02, F03, and F04), thermal management technologies (F22, F23, F24, F25 and F28), optical technologies (B41, G02, G03), construction technologies (E01, E02 and E21), and paper processing (B31 and D21. The partition of these cohesive cores agrees with the common knowledge of the relatedness and distinctions of corresponding technologies.

Our optimal partition also reveals a more meaningful structure than the community structure identified by the Louvain method, which shows 5 large communities and the largest community with 41 vertices. Each of the 5 large communities encapsulates several cores together with a few peripheral technologies that are differentiable and identified by our method (Table B2). For example, one large community (#1 in Table B2) encapsulates several cores, such as thermal management (F22, F23, F24, F25 and F28), biomedical (A01, C12, A61, and C07), chemical processing (C01, C02, C10, and B01), food processing (A21 and A23), and solid material processing (B03 and B07). Within the community, these distinct types of technology are not further discerned. This community also contains a few technologies that are in the periphery as identified in our partition, such as tobacco (A24), grain milling (B02), sugar production (C13), combinatorial chemistry (C40), and nuclear technology (G21). Such technologies are generally peripheral by nature, but are not discerned from those highly connected technologies within the same community.

Therefore, our multicores-periphery partition is more meaningful than the single core-periphery and community partitions in revealing the innate relationships of different technologies in the network[1]. Such benefits are achieved by effectively discerning various cohesive groups of highly related technologies (i.e., cores), as well as discerning dense cohesive groups (i.e., cores) from the isolated insignificant technologies (i.e., periphery) that are only loosely connected to the rest of the technology space.

---

[1] Note that, in this specific case of the technology network, the more meaningful partition turns out to be finer than the others in comparison, because the additional nuances are valuable. However, in other networks, the most meaningful partition may be coarse. In theory, the fineness or coarsens of the most meaningful partition depends on the innate structure of the examined network and emerges in the computation result. In brief, our algorithm is not aimed to deliver a finer or coarser division of a network, but find the partition that most meaningfully reveals the innate structure of the examined network.

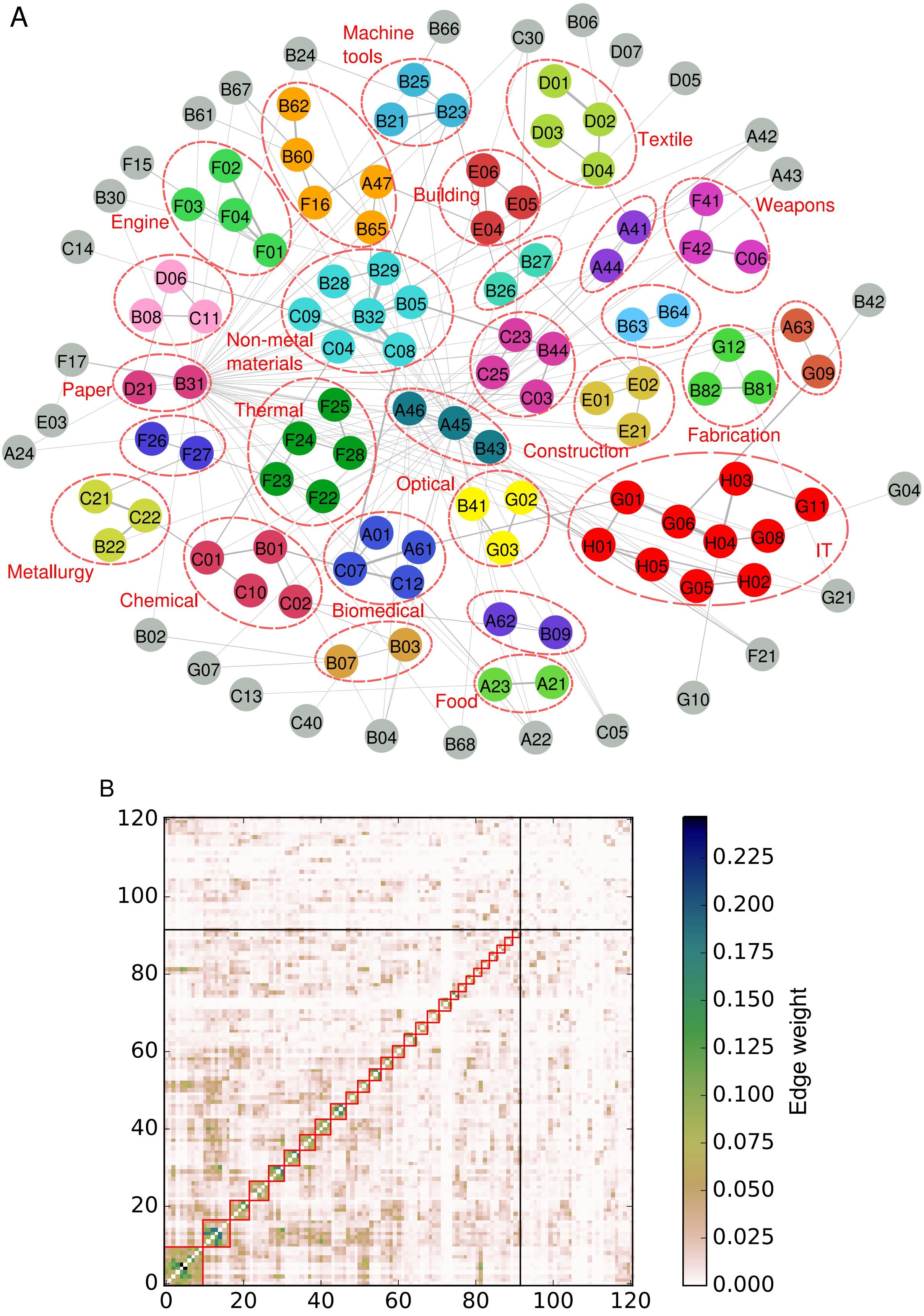

Fig. 4. Multicores-periphery structure of (A) the patent technology network, in which only the strongest edges of the original network are visualized so that the number of edges is twice the number of vertices, for a clear visualization as suggested by Hidalgo et al. (2007), and (B) its matrix representation. Each red circle represents a core, and the vertex colors denote the cores; the peripheral vertices are in grey.

## 5 Conclusion

Although many real-world networks exhibit a *multicores-periphery* structure, neither a formal definition nor quantitative methods to directly identify the structure exist. In this paper, we first formally defined the *multicores-periphery* structure in networks, i.e., multiple dense cores and a sparse periphery. The definition is distinguished from the single core-periphery structure, or multiple peripheries attached to multiple cores. This definition also does not require cores to be rigid cliques of pre-determined local densities. According to the definition, we have further introduced a method to directly detect the optimal multicores-periphery structure of a network.

Previous core-periphery algorithms focused on detecting a general single core and periphery, whereas previous community detection algorithms did not discern a periphery from dense cohesive communities. Although the studies that defined cores as cliques of maximum or pre-determined local densities may detect multiple dense cores, they did not consider the global network structure and the relative distinction from the periphery. Our detection method is dedicated to directly detecting the multicores-periphery structure. It does not pre-require a pre-determined local density of the cores (as for clique or k-core detection), but statistically identifies the optimal partition that maximizes the distinction between the density in cores and the density outside cores (i.e., the periphery).

We have illustrated the differences between our proposed algorithm and two state-of-the-art algorithms for detecting core-periphery structure and community structure via the applications to two distinct real-world networks. The multicores-periphery structures identified by our algorithm in these two real-world networks provide more systematic, nuanced and meaningful characterizations of these networks than traditional single core-periphery and community structures. The new definition and detection method for *multicores-periphery structure* may enable new analyses and understanding of many real-world biological, ecological, social and technological networks, which are best characterized by a multicores-periphery structure.

One area of future work is to apply our algorithm to analyze the dynamics of multicores-periphery networks (Csermely, 2018; Liu et al., 2015; Pan et al., 2012), whereas we only focus on the structural aspect of networks in this paper. Another valuable future research direction is to extend our algorithm for the analysis of unweighted, directed and signed networks. In addition, it will be worthwhile to further revise the algorithm or develop new algorithms with improved computation efficiency and speed for the analysis of multicores-periphery structures in very large-scale complex networks, in contrast to the two small networks in the present paper.

We hope the readers view this paper as a beginning and invitation for future research and development of more efficient algorithms to directly detect the multicores-periphery structure. The methodological development for the multicores-periphery structure is clearly nascent, compared to the extensive methods that have been developed for detecting communities, local cliques or *k*-cores, and single core-periphery dichotomy. Together with methodological development, we also hope more network analyses will take the lens of the multicores-periphery structure to understand, design and manage large-scale complex networked systems in diverse contexts and domains.

## Acknowledgements

This work was supported by the SUTD-MIT International Design Centre and the Singapore Ministry of Education Tier 2 Academic Research Grants.

## Appendix A: Analysis of Zachary's Karate Club Network

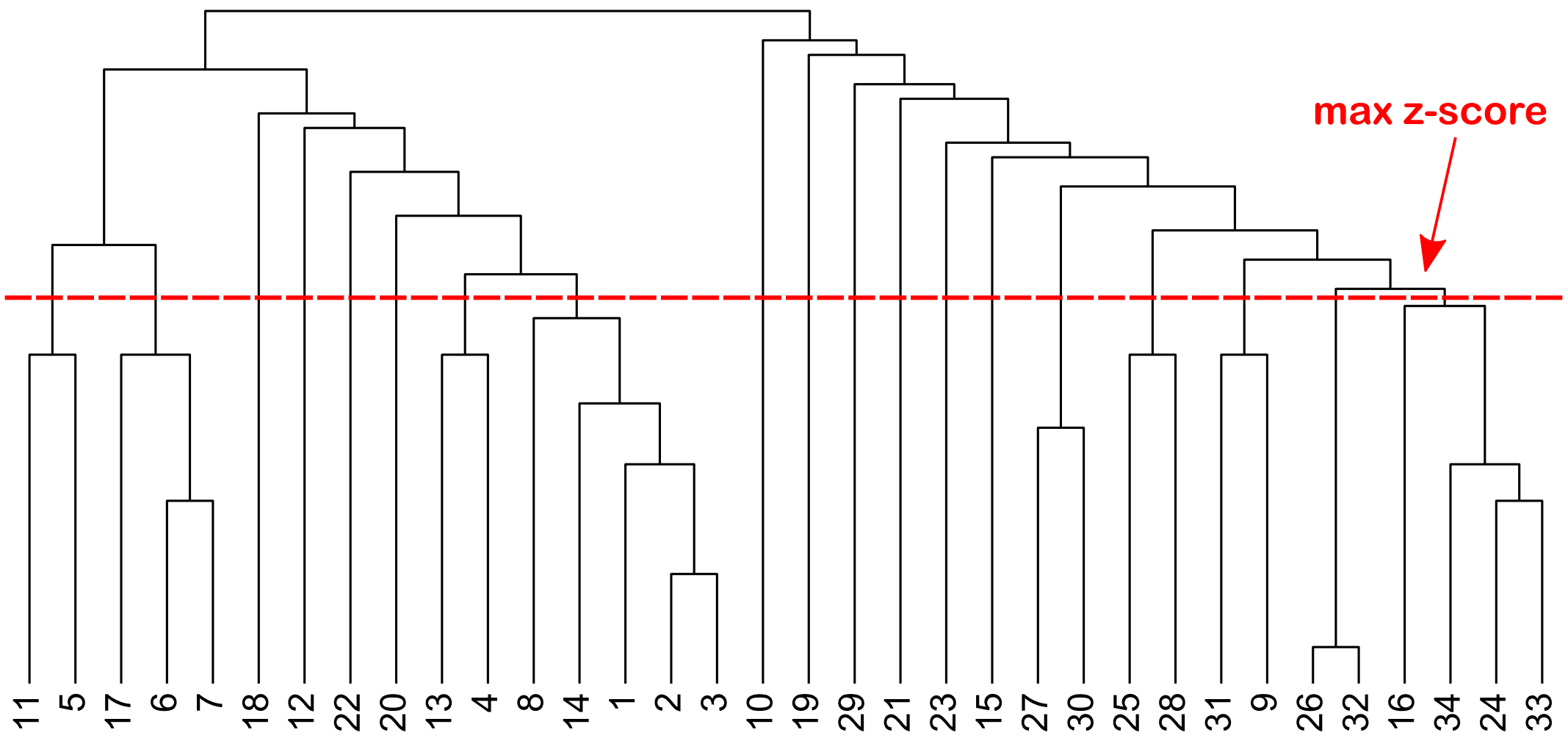

Fig. A1. The dendrogram of Zachary's karate club network.

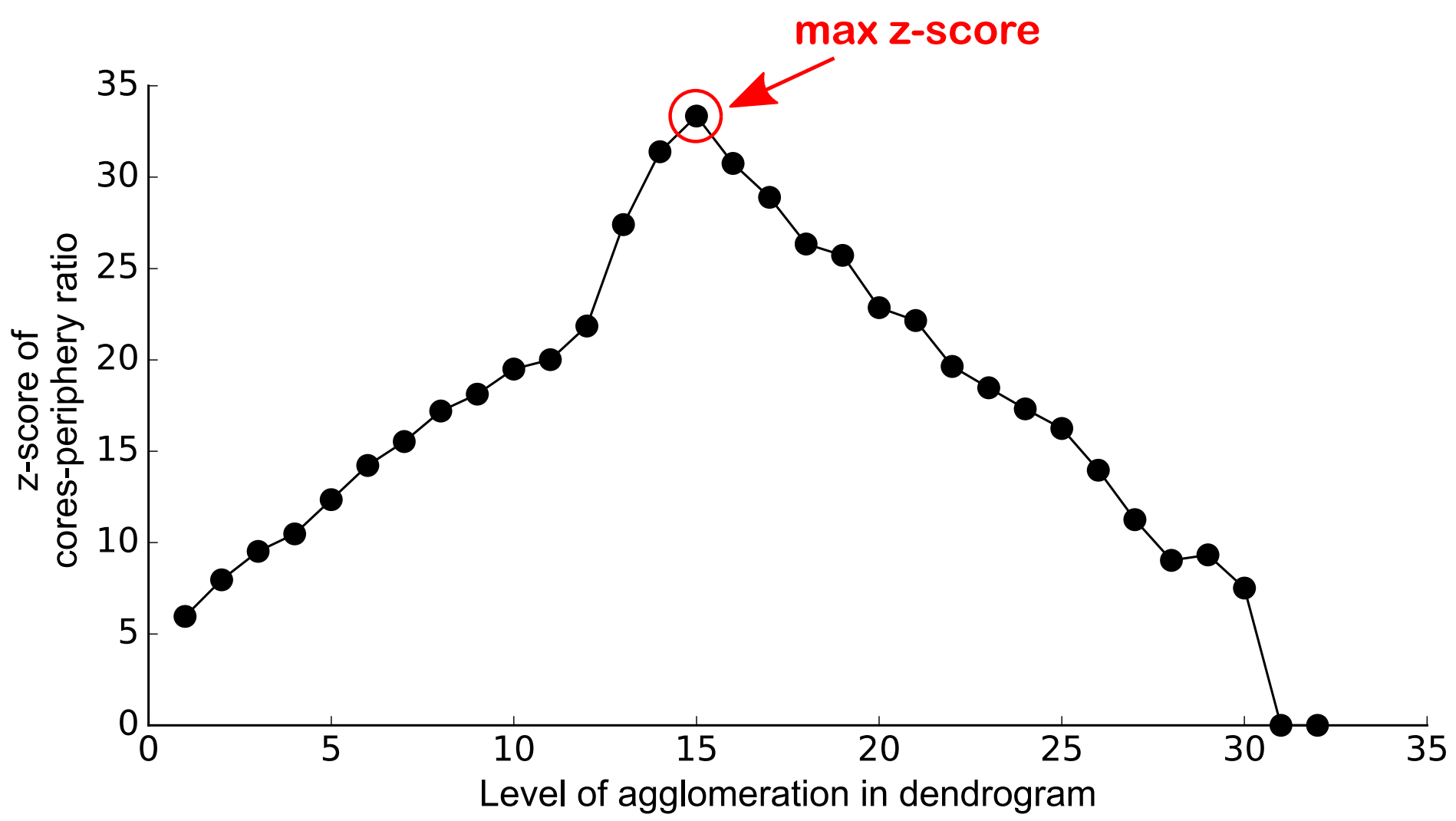

Fig. A2. $z$-score curve for deciding the optimal partition in Zachary's karate club network. For each partition, $z$-score is calculated comparing the empirical network with an ensemble of one million randomized networks.

Table A1. Partition of 34 vertices in the Zachary's karate club network

| Vertex ID (Persons) | Core ID or Periphery (P) | Core (C) or Periphery (P) | Community ID |
|---|---|---|---|
| 8 | 1 | P | 1 |
| 14 | 1 | C | 1 |
| 1 | 1 | C | 1 |
| 2 | 1 | C | 1 |
| 3 | 1 | C | 1 |
| 16 | 2 | P | 3 |
| 34 | 2 | C | 3 |
| 24 | 2 | P | 2 |
| 33 | 2 | C | 3 |
| 17 | 3 | P | 4 |
| 6 | 3 | P | 4 |
| 7 | 3 | P | 4 |
| 26 | 4 | P | 2 |
| 32 | 4 | P | 2 |
| 27 | 5 | P | 3 |
| 30 | 5 | P | 3 |
| 11 | 6 | P | 4 |
| 5 | 6 | P | 4 |
| 13 | 7 | P | 1 |
| 4 | 7 | P | 1 |
| 25 | 8 | P | 2 |
| 28 | 8 | P | 2 |
| 31 | 9 | P | 3 |
| 9 | 9 | C | 3 |
| 10 | P | P | 3 |
| 12 | P | P | 1 |
| 15 | P | P | 3 |
| 18 | P | P | 1 |
| 19 | P | P | 3 |
| 20 | P | P | 1 |
| 21 | P | P | 3 |
| 22 | P | P | 1 |
| 23 | P | P | 3 |
| 29 | P | P | 2 |

"Core ID or periphery (P)": results obtained by the method proposed in the paper.

"Core (C) or periphery (P)": results obtained by Borgatti and Everett's method.

"Community ID": results obtained by the Louvain method.

## Appendix B: Analysis of Patent Technology Network

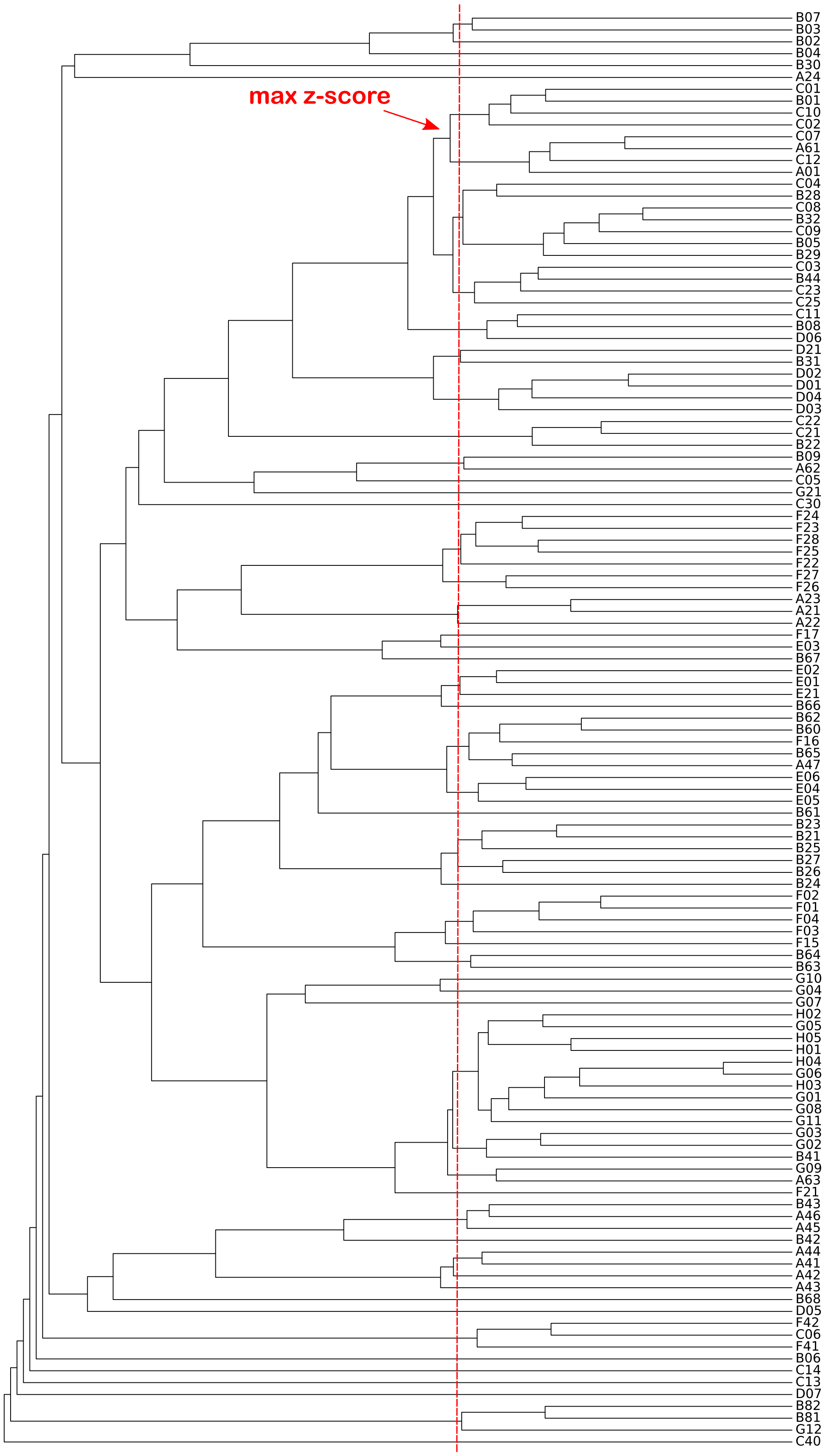

Fig. B1. The dendrogram of patent technology network.

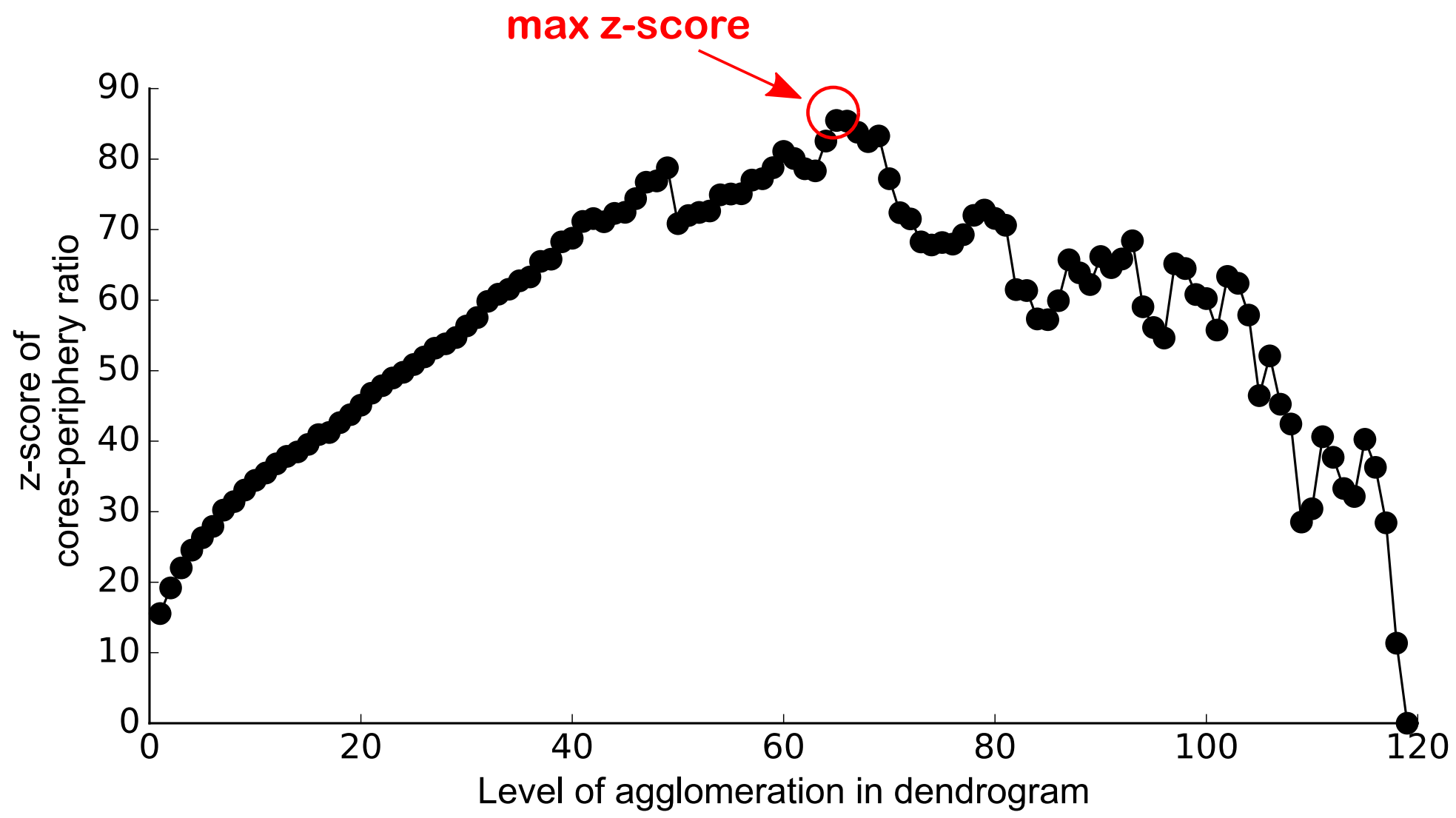

Fig. B2. $z$-score curve for deciding the optimal partition in patent technology network. For each partition, $z$-score is calculated comparing the empirical network with an ensemble of one million randomized networks.

Table S2. Partitions of 121 vertices in the technology network

| Vertex ID (IPC3 Class) | Description | Core ID or Periphery (P) | Core (C) or Periphery (P) | Community ID |
|---|---|---|---|---|
| G11 | Information Storage | 1 | C | 3 |
| G08 | Signaling | 1 | C | 3 |
| G01 | Measuring & Testing | 1 | C | 3 |
| H03 | Electronic Circuitry | 1 | C | 3 |
| G06 | Computing | 1 | C | 3 |
| H04 | Electric Communication | 1 | C | 3 |
| H01 | Electric Elements | 1 | C | 3 |
| H05 | Electric Techniques | 1 | C | 3 |
| G05 | Controlling & Regulating | 1 | C | 3 |
| H02 | Electric Power | 1 | C | 3 |
| B29 | Plastics-Working | 2 | C | 4 |
| B05 | Spraying or Atomizing | 2 | C | 4 |
| C09 | Organic Material Applications | 2 | C | 4 |
| B32 | Layered Products | 2 | C | 4 |
| C08 | Organic Macromolecular Compounds | 2 | C | 4 |
| B28 | Cement, Clay or Stone Working | 2 | C | 4 |
| C04 | Building Materials | 2 | C | 4 |
| A47 | Furniture & Appliances | 3 | C | 2 |
| B65 | Filamentary Material Handling | 3 | C | 2 |
| F16 | Machine Elements | 3 | C | 2 |
| B60 | Vehicles in General | 3 | C | 2 |
| B62 | Land Vehicles | 3 | C | 2 |
| F22 | Steam Generation | 4 | C | 1 |
| F25 | Refrigeration, Liquefaction or Solidification | 4 | C | 1 |
| F28 | Heat Exchange in General | 4 | C | 1 |
| F23 | Combustion Apparatus & Processes | 4 | C | 1 |
| F24 | Heating & Ventilating | 4 | C | 1 |
| A01 | Agriculture | 5 | C | 1 |
| C12 | Biochemistry & Genetic Engineering | 5 | C | 1 |
| A61 | Medical & Hygiene | 5 | C | 1 |
| C07 | Organic Chemistry | 5 | C | 1 |
| D03 | Weaving | 6 | C | 4 |
| D04 | Braiding & Knitting | 6 | C | 4 |
| D01 | Threads or Fibers | 6 | C | 4 |
| D02 | Yarns or Rope Finishing | 6 | C | 4 |
| C02 | Water Treatment | 7 | C | 1 |
| C10 | Fuels & Lubricants | 7 | C | 1 |
| B01 | Physical or Chemical Processes | 7 | C | 1 |
| C01 | Inorganic Chemistry | 7 | C | 1 |
| C25 | Electrolysis or Electrophoresis | 8 | C | 4 |
| C23 | Coating Metallic Material | 8 | C | 4 |
| B44 | Decorative Arts | 8 | C | 4 |
| C03 | Glass, Mineral or Wool | 8 | C | 4 |
| F03 | Machines or Engines for Liquids | 9 | C | 2 |
| F04 | Pumps | 9 | C | 2 |
| F01 | Machines or Engines in General | 9 | C | 2 |
| F02 | Combustion Engines | 9 | C | 2 |
| D06 | Textile Treatment | 10 | C | 4 |
| B08 | Cleaning | 10 | C | 4 |
| C11 | Fat, Oil & Wax Processing | 10 | C | 4 |
| B41 | Printing | 11 | C | 3 |
| G02 | Optics | 11 | C | 3 |
| G03 | Photography, Electrograph & Holography | 11 | C | 3 |
| B22 | Casting & Metallurgy | 12 | C | 4 |
| C21 | Metallurgy of Iron | 12 | C | 4 |
| C22 | Metallurgy of Non-ferrous Metals or Alloys | 12 | C | 4 |
| B25 | Workshop Equipment | 13 | C | 2 |
| B21 | Mechanical Metal-Working | 13 | C | 2 |
| B23 | Machine Tools | 13 | C | 2 |
| E05 | Locks, Keys & Safes | 14 | C | 2 |
| E04 | Building Construction | 14 | C | 2 |
| E06 | Building & Vehicle Closures & Ladders | 14 | C | 2 |
| F41 | Weapons | 15 | C | 2 |
| C06 | Explosives & Matches | 15 | C | 2 |
| F42 | Ammunition & Blasting | 15 | C | 2 |
| A45 | Hand or Travelling Articles | 16 | C | 2 |
| A46 | Brushware | 16 | C | 2 |
| B43 | Writing & Drawing Implements | 16 | C | 2 |
| E21 | Drilling & Mining | 17 | C | 2 |
| E01 | Road, Railway & Bridge Construction | 17 | C | 2 |
| E02 | Hydraulic & Construction Engineering | 17 | C | 2 |
| G12 | Instrument Details | 18 | P | 5 |
| B81 | Micro-Structural Technology | 18 | P | 5 |
| B82 | Nano-Technology | 18 | P | 5 |
| A21 | Baking | 19 | C | 1 |
| A23 | Food Processing | 19 | C | 1 |
| F26 | Drying | 20 | C | 1 |
| F27 | Furnaces, Kilns & Ovens | 20 | C | 1 |
| B26 | Hand Cutting Tools | 21 | C | 2 |
| B27 | Wood-Working | 21 | C | 2 |
| A63 | Sports & Amusements | 22 | C | 2 |
| G09 | Info graphics & Display | 22 | C | 3 |
| A41 | Clothing | 23 | C | 2 |
| A44 | Haberdashery & Jewelry | 23 | C | 2 |
| B03 | Solid Material Separation | 24 | C | 1 |
| B07 | Separation of Solids | 24 | C | 1 |
| B63 | Ships | 25 | C | 2 |
| B64 | Aircraft | 25 | C | 2 |
| A62 | Life-saving | 26 | C | 1 |
| B09 | Disposal of Waste | 26 | C | 1 |
| B31 | Paper Articles | 27 | C | 4 |
| D21 | Paper & Cellulose Making | 27 | C | 4 |
| A22 | Butchering | P | C | 1 |
| A24 | Tobacco | P | C | 1 |
| A42 | Headwear | P | C | 2 |
| A43 | Footwear | P | C | 2 |
| B02 | Grain Milling | P | C | 1 |
| B04 | Centrifugal Machines | P | C | 1 |
| B06 | Mechanical Vibration | P | C | 4 |
| B24 | Grinding & Polishing | P | C | 2 |
| B30 | Press Machine | P | C | 4 |
| B42 | Sheet-binding | P | C | 2 |
| B61 | Railways | P | C | 2 |
| B66 | Hoisting & Hauling Machines | P | C | 2 |
| B67 | Liquid Containers | P | C | 2 |
| B68 | Saddlery & Upholstery | P | C | 2 |
| C05 | Fertilizers | P | C | 1 |
| C13 | Sugar Production | P | P | 1 |
| C14 | Leather | P | P | 4 |
| C30 | Crystal Growth | P | C | 4 |
| C40 | Combinatorial Chemistry | P | P | 1 |
| D05 | Sewing | P | C | 4 |
| D07 | Ropes or Cables in General | P | P | 4 |
| E03 | Water Supply & Sewerage | P | C | 2 |
| F15 | Hydraulics & Pneumatics | P | C | 2 |
| F17 | Storing or Distributing of Liquids | P | C | 1 |
| F21 | Lighting | P | C | 3 |
| G04 | Horology | P | C | 3 |
| G07 | Checking-devices | P | C | 3 |
| G10 | Musical Instruments & Acoustics | P | C | 3 |
| G21 | Nuclear Technology | P | C | 1 |